\documentclass[final,3p,11pt]{elsarticle}

\usepackage{url,hyperref,lineno,microtype,subcaption}

\usepackage{amssymb}

\usepackage{lineno}

\usepackage{tabularx}
\usepackage{booktabs}

\journal{First Monday}

\begin{document}

\begin{frontmatter}



\title{Should ChatGPT be Biased?\\ Challenges and Risks of Bias in Large Language Models}


\author[inst1,inst2]{Emilio Ferrara}

\affiliation[inst1]{organization={Thomas Lord Department of Computer Science},
            addressline={University of Southern California}, 
            city={Los Angeles},
            postcode={90007}, 
            state={CA},
            country={USA}}
\affiliation[inst2]{organization={USC Information Sciences Institute},
            addressline={4676 Admiralty Way \#1001}, 
            city={Marina del Rey},
            postcode={90292}, 
            state={CA},
            country={USA},
            email={\\ E-mail: emiliofe@usc.edu}}



\begin{abstract}

As generative language models, exemplified by ChatGPT, continue to advance in their capabilities, the spotlight on biases inherent in these models intensifies. This article delves into the distinctive challenges and risks associated with biases specifically in large-scale language models. We explore the origins of biases, stemming from factors such as training data, model specifications, algorithmic constraints, product design, and policy decisions. Our examination extends to the ethical implications arising from the unintended consequences of biased model outputs. In addition, we analyze the intricacies of mitigating biases, acknowledging the inevitable persistence of some biases, and consider the consequences of deploying these models across diverse applications, including virtual assistants, content generation, and chatbots. Finally, we provide an overview of current approaches for identifying, quantifying, and mitigating biases in language models, underscoring the need for a collaborative, multidisciplinary effort to craft AI systems that embody equity, transparency, and responsibility. This article aims to catalyze a thoughtful discourse within the AI community, prompting researchers and developers to consider the unique role of biases in the domain of generative language models and the ongoing quest for ethical AI.

\end{abstract}



\begin{keyword}
Artificial Intelligence \sep Generative AI \sep Bias \sep Large Language Models \sep ChatGPT \sep GPT-4
\end{keyword}

\end{frontmatter}



\hypertarget{introduction}{%
\section{Introduction}\label{introduction}}


In recent years, there has been a remarkable surge in the realm of artificial intelligence (AI), marked by the emergence of transformative technologies like ChatGPT and other generative language models (\cite{transformer, bert, gpt, gpt2, gpt3, ppo, rlhf, instructgpt, gpt4}). These AI systems represent a class of sophisticated models designed to excel in generating human-like text and comprehending natural language. Using deep learning techniques and vast datasets, they have the ability to discern intricate patterns, make contextual inferences, and generate coherent and contextually relevant responses to a diverse range of inputs (\cite{neural-probabilistic-language-models, word2vec, seq2seq, transformer, bert}). By mirroring human language capabilities, these models have unveiled a multitude of applications, ranging from chatbots and virtual assistants to translation services and content generation tools (\cite{young2018recent}).

Language models have ushered in a transformative era, underpinning the development of chatbots that emulate human interaction in conversations (\cite{chen2017survey}). These chatbots have become vital tools, simplifying customer service, technical support, and information queries with human-like interaction. Furthermore, the integration of language models into virtual assistants has endowed them with the ability to provide precise and contextually appropriate responses to user inquiries (\cite{zhang2018personalizing}). Virtual assistants, thus enhanced, become indispensable aides, capable of managing tasks ranging from appointment scheduling to web searches and smart home device control (\cite{wen2016network}).

In the sphere of translation, harnessing the prowess of large language models facilitates markedly improved and fluent translations that span multiple languages, including those with limited resources (\cite{wang2019learning, karakanta2018neural, pourdamghani2019neighbors, costa2022no, ranathunga2023neural}). Such capabilities not only foster enhanced cross-linguistic communication but can also enable timely solutions during emergencies and crises, especially in regions where low-resource languages or indigenous dialects are spoken (\cite{christianson2018overview}).

Moreover, the aptitude of language models to generate coherent and contextually pertinent text has rendered them invaluable in the realm of content creation. Acknowledged for their proficiency in producing various types of content, spanning articles, social media posts, and marketing materials, they have established a profound impact (\cite{yang2022botometer, ferrara2023social}).

These applications, among many others, underscore the transformative prowess of generative language models across an array of industries and sectors. However, as their adoption proliferates and their influence extends into ever-diverse domains (\cite{gilson2023does, choi2023chatgpt}), it is imperative to confront the distinctive challenges posed by the potential biases that may be established within these models. These biases can have profound implications for users and society at large, highlighting the urgent need for comprehensive examination and mitigation of these issues \cite{ferrara2023genai}.


\begin{table}[t!]
\centering \footnotesize
\begin{tabular}{|l|p{7.4cm}|p{4cm}|}
\hline
\textbf{Contributing Factor} & \multicolumn{1}{c|}{\textbf{Description}}                             & \multicolumn{1}{c|}{\textbf{References}} \\
\hline
Training Data                         & Biases in the source material or the selection process for training data can be absorbed by the model and reflected in its behavior.                            & \cite{bolukbasi2016man, caliskan2017semantics, blodgett2020language, parrots}    \\
\hline
Algorithms                            & Biases can be introduced or amplified through algorithms that place more importance on certain features or data points.                                       & \cite{solaiman2019release, blodgett2020language, hovy2021five}                   \\
\hline
Labeling and Annotation               & In (semi)supervised learning scenarios, biases may emerge from subjective judgments of human annotators providing labels or annotations for the training data.
& \cite{munro2010crowdsourcing, buolamwini2018gender, bender2018data}              \\
\hline
Product Design Decisions              & Biases can arise from prioritizing certain use cases or designing user interfaces for specific demographics or industries, inadvertently reinforcing existing biases and excluding different perspectives.                                                & \cite{kleinberg2016inherent, benjamin2020race}                                   \\
\hline
Policy Decisions                      & Developers might implement policies that prevent (or encourage) a given model behavior. For example, guardrails that modulate the behavior of ChatGPT and Bing-AI were designed to mitigate unintended toxic model behaviors or prevent malicious abuse.               & \cite{doshi2017towards, binns2018fairness, crawford2019ai, prates2020assessing}  \\
\hline
\end{tabular}
\caption{Factors Contributing to Bias in AI Models}
\label{table:factors_contributing_to_bias}
\end{table}

\section{Defining bias in generative language models}

\subsection{Factors contributing to bias in Large Language Models}

Bias, in the context of large language models such as GPT-4 (\cite{gpt4, gpt4card, bubeck2023sparks}) and predecessors (\cite{gpt, gpt2, gpt3}), or other state-of-the-art alternatives (\cite{t5, llama}; including multimodal variants, \cite{visualchatgpt}), can be defined as the presence of systematic misrepresentations, attribution errors, or factual distortions that result in favoring certain groups or ideas, perpetuating stereotypes, or making incorrect assumptions based on learned patterns. Biases in such models can arise due to several factors (\textit{cf.}, Table \ref{table:factors_contributing_to_bias}).

One factor is the \textit{training data}. If the data used to train a language model contain biases, either from the source material or through the selection process, these biases can be absorbed by the model and subsequently reflected in its behavior (\cite{bolukbasi2016man, caliskan2017semantics, blodgett2020language, parrots}).
Biases can also be introduced through the \textit{algorithms} used to process and learn from the data. For example, if an algorithm places more importance on certain features or data points, it may unintentionally introduce or amplify biases present in the data (\cite{solaiman2019release, blodgett2020language, hovy2021five}).
In (semi)supervised learning scenarios, where \textit{human annotators} provide labels or annotations for the training data, biases may emerge from the subjective judgments of the annotators themselves, influencing the model's understanding of the data (\cite{munro2010crowdsourcing, buolamwini2018gender, bender2018data}).

The choice of which \textit{use cases} to prioritize or the design of user interfaces can also contribute to biases in large language models. For example, if a language model is primarily designed to generate content for a certain demographic or industry, it may inadvertently reinforce existing biases and exclude different perspectives (\cite{kleinberg2016inherent, benjamin2020race}).
Lastly, \textit{policy decisions} can play a role in the manifestation of biases in language models. The developers of both commercial and openly available language models might implement policies that prevent (or encourage) a given model behavior. For example, both OpenAI and Microsoft have deliberate guardrails that modulate the behavior of ChatGPT and Bing-AI to mitigate unintended toxic model behaviors or prevent malicious abuse (\cite{doshi2017towards, binns2018fairness, crawford2019ai, prates2020assessing}).

\subsection{Types of biases in Large Language Models}

Large language models, which are commonly trained from vast amounts of text data present on the Internet, inevitably absorb the biases present in such data sources. These biases can take various forms (\textit{cf.}, Table \ref{table:types_of_biases}).

Demographic biases arise when the training data over-represents or under-represents certain demographic groups, leading the model to exhibit biased behavior towards specific genders, races, ethnicities, or other social groups (\cite{munro2010crowdsourcing, bolukbasi2016man, caliskan2017semantics, buolamwini2018gender, parrots, kirk2021bias}).
Cultural biases occur when large language models learn and perpetuate cultural stereotypes or biases, as they are often present in the data used for training. This can result in the model producing outputs that reinforce or exacerbate existing cultural prejudices (\cite{bordia2019identifying, ribeiro2020beyond, blodgett2020language}).
Linguistic biases emerge since the majority of the internet's content is in English or a few other dominant languages, making large language models more proficient in these languages. This can lead to biased performance and a lack of support for low-resource languages or minority dialects (\cite{conneau2017word, johnson2017google, pires2019multilingual, ruder2019survey, parrots}).
Temporal biases appear as the training data for these models are typically restricted to limited time periods or have temporal cutoffs. This may cause the model to be biased when reporting on current events, trends, and opinions. Similarly, the model's understanding of historical contexts or outdated information may be limited due to a lack of temporally representative data (\cite{gpt, zellers2019defending, mccoy2019right, smith2020contextual}).
Confirmation biases in the training data may result from individuals seeking out information that aligns with their pre-existing beliefs. Consequently, large language models may inadvertently reinforce these biases by providing outputs that confirm or support specific viewpoints (\cite{bolukbasi2016man, caliskan2017semantics, bert, modelsheets}).
Lastly, ideological and political biases can be learned and propagated by large language models due to the presence of such biases in their training data. This can lead to the model generating outputs that favor certain political perspectives or ideologies, thereby amplifying existing biases (\cite{garg2018word, dixon2018measuring, mccoy2019right, mcgee2023chat}).


This paper aims to explore the question of whether language models like
GPT-4 (\cite{gpt4, gpt4card, bubeck2023sparks}), its prior versions (\cite{gpt, gpt2, gpt3}), or other commercial or open-source
alternatives (\cite{t5, llama, palm}) that power applications like ChatGPT (\cite{instructgpt}) (or similar) should be
biased or unbiased, taking into account the implications and risks of both
perspectives. By examining the ethical, practical, and societal
consequences of each viewpoint, we hope to contribute to the ongoing
discussion surrounding responsible language model development and use.
Through this exploration, our goal is to provide insights that can help
guide the future evolution of GPT-style and other generative language models
toward more ethical, fair, and beneficial outcomes while minimizing
potential harm.

\begin{table}[t!]
\centering \footnotesize
\begin{tabular}{|l|p{7cm}|p{4cm}|}
\hline
\textbf{Types of Bias} & \multicolumn{1}{c|}{\textbf{Description}}                                                 & \multicolumn{1}{c|}{\textbf{References}} \\
\hline
Demographic Biases     & These biases arise when the training data over-represents or under-represents certain demographic groups, leading the model to exhibit biased behavior towards specific genders, races, ethnicities, or other social groups.                                                                                        & \cite{munro2010crowdsourcing, bolukbasi2016man, caliskan2017semantics, buolamwini2018gender, parrots, kirk2021bias} \\
\hline
Cultural Biases        & Large language models may learn and perpetuate cultural stereotypes or biases, as they are often present in the data used for training. This can result in the model producing outputs that reinforce or exacerbate existing cultural prejudices.                    & \cite{bordia2019identifying, ribeiro2020beyond, blodgett2020language} \\
\hline
Linguistic Biases      & Since the majority of the internet's content is in English or a few other dominant languages, large language models tend to be more proficient in these languages. This can lead to biased performance and a lack of support for low-resource languages or minority dialects.                             
& \cite{conneau2017word, johnson2017google, pires2019multilingual, ruder2019survey, parrots} \\
\hline
Temporal Biases        & The training data for these models are typically restricted to limited time periods, or have temporal cutoffs, which may cause the model to be biased when reporting on current events, trends, and opinions. Similarly, the model's understanding of historical contexts or outdated information may be limited for lack of temporally representative data.                                                     & \cite{gpt, zellers2019defending, mccoy2019right, smith2020contextual} \\
\hline
Confirmation Biases    & The training data may contain biases that result from individuals seeking out information that aligns with their pre-existing beliefs. Consequently, large language models may inadvertently reinforce these biases by providing outputs that confirm or support specific viewpoints.                       & \cite{bolukbasi2016man, caliskan2017semantics, bert, modelsheets} \\
\hline
Ideological \& Political Biases & Large language models can also learn and propagate the political and ideological biases present in their training data. This can lead to the model generating outputs that favor certain political perspectives or ideologies, thereby amplifying existing biases.
& \cite{garg2018word, dixon2018measuring, mccoy2019right, mcgee2023chat} \\
\hline
\end{tabular}
\caption{Types of Biases in Large Language Models}
\label{table:types_of_biases}
\end{table}

\hypertarget{background}{%
\section{Why are generative language models prone to bias?}\label{background}}

\subsection{Biases from the data}
ChatGPT and other applications based on large language models are
trained using a process that primarily relies on unsupervised learning,
a machine learning technique that enables models to learn patterns and
structures from vast amounts of unlabelled data (\cite{jiang2020can, carlini2021extracting}). In most cases with these
language models, the data consists of extensive text corpora available
on the internet, which includes websites, articles, books, and other
forms of written content (\cite{bert, t5, llama, palm}).

ChatGPT in particular is trained on a diverse range of internet text
datasets that encompass various domains, genres, and languages. While
the specifics of the dataset used for GPT-4 are proprietary
(\cite{gpt4, gpt4card}), the data sources utilized for training its
predecessor, GPT-3, likely share similarities. For GPT-3 and
predecessors, the primary dataset used was WebText (\cite{gpt}), which is
an ever-growing large-scale collection of web pages (\cite{gpt2, gpt3}).
WebText was created by crawling the internet and gathering text from web
pages. The sources of data include, but are not limited to:

\begin{itemize}
    \item 
\textbf{Websites}: Text is extracted from a wide array of websites,
covering topics such as news, blogs, forums, and informational websites
like Wikipedia. This enables the model to learn from diverse sources and
gain knowledge on various subjects. 
    \item
\textbf{Books}: Text from books available online, including both fiction
and non-fiction, contributes to the training data. This helps the model
to learn different writing styles, narrative structures, and a wealth of
knowledge from various fields.
    \item
\textbf{Social media platforms}: Content from social media platforms,
like Twitter, Facebook, and Reddit, is incorporated to expose the model
to colloquial language, slang, and contemporary topics of discussion.
    \item
\textbf{Conversational data}: To improve the model's conversational
abilities, text from chat logs, comment sections, and other
conversational sources are also included in the training dataset.
\end{itemize}

The developers of ChatGPT note that the WebText data is preprocessed and
filtered to remove low-quality content, explicit material, web and social spam \cite{ferrara2019history, ferrara2022twitter}, and other
undesirable text before being fed into the model (\cite{gpt2, gpt3}). 
However, due to the
vast scale of the data and the limitations of current filtering
techniques, some undesirable or biased content may still seep into the
training dataset, affecting the behavior of the resulting model.
In addition to WebText, GPT-3 was further trained using a filtered version of the Common Crawl dataset (\url{https://commoncrawl.org}), a publicly available, massive web-crawled dataset that contains raw web page data, extracted metadata, and text content from billions of web pages in multiple languages (\cite{gpt3}).

Another commonly-used dataset for language model training is \emph{The
Pile} (\cite{piledataset, piledatasheet}) an extensive and
diverse collection of 22 smaller datasets, combining various sources of
scientific articles, books, and web content. It is designed for training
large-scale language models, particularly in the domain of scientific
research and understanding.

\subsection{Biases from the models}

During the training process, generative language models are exposed to billions of sentences and phrases, allowing them to learn the intricate relationships between words, grammar, context, and meaning (\cite{jiang2020can, carlini2021extracting}). As they process the text data, they gradually acquire natural language generation capabilities, enabling them to produce coherent and contextually relevant responses to various inputs. However, some capabilities of these models can lead to bias (\textit{cf.}, Table \ref{tab:biases_from_models}):

\paragraph{Generalization} 
One crucial aspect of these models is their ability to generalize, which allows them to apply the knowledge gained from their training data to new and previously unseen inputs, providing contextually relevant responses and predictions even in unfamiliar situations. However, this ability also raises concerns about potential biases, as models may inadvertently learn and perpetuate biases present in their training data, even if the data has been filtered and cleaned to the extent possible (\cite{gururangan2018annotation, caliskan2017semantics}).

\paragraph{Propagation}
As these models learn from the patterns and structures present in their training data, they may inadvertently absorb and propagate biases they encounter, such as adopting stereotypes, favoring certain groups or ideas, or making assumptions based on learned patterns that do not accurately represent the full spectrum of human experience. This propagation of biases during training poses significant challenges to the development of fair and equitable AI systems, as biased models can lead to unfair treatment, reinforce stereotypes, and marginalize certain groups (\cite{bolukbasi2016man, dev2019attenuating, ferrara2023butterfly}).

\paragraph{Emergence}
In large language models, the phenomenon of emergence, which refers to the spontaneous appearance of unanticipated capabilities despite these functionalities not being explicitly encoded within the model's architecture or training data, can also result in unexpected biases due to the intricate interplay between model parameters and biased training data (\cite{chainofthought, emergence}). The high-dimensional representations and non-linear interactions in these models make it difficult to predict or control these emergent biases, which may manifest in various ways, such as stereotyping, offensive language, or misinformation. To address this challenge, researchers are exploring bias mitigation strategies during training, fine-tuning with curated datasets, and post-hoc emergent bias analyses (\cite{big-bench, wei2022emergent}).

\paragraph{Non-linearity}
The non-linear relationships between biases in the system or data and their real-world impact imply that small biases may have massive negative effects, and large biases might not result in significant consequences. This disproportionality arises due to the complex interdependencies between the model parameters and the high-dimensional representations learned during training \cite{ferrara2023butterfly}. Randomized controlled trials could be used to draw causal relationships between the extent of each bias and their effects. In the absence of that, due to ethical reasons, multifaceted approaches involving in-depth analysis of model behavior, rigorous evaluation with diverse benchmarks, and the application of mitigation techniques that account for the nonlinear nature of emergent biases are needed \cite{chiappa2019path}.

\paragraph{Alignment}
To address these issues, a strategy known as Reinforcement Learning with Human Feedback (RLHF) (\cite{rlhf}) was developed to fine-tune large language models like ChatGPT to reduce their biases and align them with human values. This approach involves collecting a dataset of human demonstrations, comparisons, and preferences to create a reward model that guides the fine-tuning process \cite{kumar2023controlled}. InstructGPT (ChatGPT's default model) (\cite{instructgpt}) is trained using RLHF and then fine-tuned using Proximal Policy Optimization (PPO), a policy optimization algorithm (\cite{ppo}). It is paramount to understand if the same principles could be exploited to deliberately misalign a model.

\begin{table}[t!]
\centering \footnotesize
\begin{tabular}{|l|p{9cm}|p{4cm}|}
\hline
\textbf{Source} & \multicolumn{1}{c|}{\textbf{Description}}           & \multicolumn{1}{c|}{\textbf{References}} \\
\hline
Generalization        & 
Models generalize knowledge from training data to new inputs, potentially leading to biased behavior if the data contains biases. This raises concerns about perpetuating biases, even if training data has been cleaned and filtered.
& \cite{gururangan2018annotation, caliskan2017semantics} \\
\hline
Propagation           & 
Models may absorb and propagate biases in training data, adopting stereotypes and favoring certain groups or ideas, or making assumptions on on non-representative learned patterns.
& \cite{bolukbasi2016man, dev2019attenuating, ferrara2023butterfly} \\
\hline
Emergence             & 
Unanticipated capabilities and biases may emerge in large language models due to complex interactions between model parameters and biased training data.  It has been proven difficult to predict or control these emergent biases.
& \cite{chainofthought, emergence, big-bench, wei2022emergent} 
\\
\hline
Non-linearity       & 
Biases in AI systems may have non-linear real-world impact, making it difficult to predict their consequences: small model biases may have massive negative effects, whereas large model biases might not cause significant consequences.  & \cite{chiappa2019path, ferrara2023butterfly} \\
\hline
Alignment & 
Reinforcement Learning with Human Feedback (RLHF) fine-tunes large language models to reduce biases and align them with human values. The same principles might be abused to lead to unfair model behaviors. & 
\cite{rlhf, instructgpt, ppo}\\
\hline

\end{tabular}
\caption{Sources of model bias in large language models and their descriptions.}
\label{tab:biases_from_models}
\end{table}

\subsection{Can bias be mitigated with human-in-the-loop approaches?}
Bias in generative language models can be mitigated to some extent with human-in-the-loop (HITL) approaches. These approaches involve incorporating human input, feedback, or oversight throughout the development and deployment of the language model, which can help address issues related to biases and other limitations. Here are some ways to integrate human-in-the-loop approaches to mitigate bias:

\begin{itemize}
    \item 

\textbf{Training data curation}: Humans can be involved in curating and annotating high-quality and diverse training data. This may include identifying and correcting biases, ensuring a balance of perspectives, and reducing the influence of controversial or offensive content (\cite{hovy2016social, parrots, hovy2021five}).
    \item
\textbf{Model fine-tuning}: Subject matter experts can guide the model fine-tuning process by providing feedback on the model's outputs, helping the model generalize better and avoid biased or incorrect responses (\cite{gururangan2020don}).
    \item
\textbf{Evaluation and feedback}: Human reviewers can evaluate the model's performance and provide feedback to developers, who can then iteratively improve the model. This feedback loop is essential for identifying and addressing bias-related issues. (\cite{modelsheets}).
    \item
\textbf{Real-time moderation}: Human moderators can monitor and review the model's outputs in real-time, intervening when necessary to correct biased or inappropriate responses. This approach can be especially useful in high-stakes or sensitive applications (\cite{park2017one}).
    \item
\textbf{Customization and control}: Users can be provided with options to customize the model's behavior, adjusting the output according to their preferences or requirements. This approach can help users mitigate bias in the model's responses by tailoring it to specific contexts or domains (\cite{gpt, bisk2020experience}).

\end{itemize}

While human-in-the-loop approaches can help mitigate bias, it is essential to recognize that they may not be able to eliminate it entirely. Bias can stem from various sources, such as the training data, fine-tuning process, or even the human reviewers themselves. However, combining machine learning techniques with human expertise can be a promising way to address some of the challenges posed by biases in generative language models.

\hypertarget{the-inevitability-of-bias}{%
\section{The inevitability of some forms of bias}\label{the-inevitability-of-bias}}
\hypertarget{bias-i-am-inevitable}{%
\subsection{Are some biases inevitable?}\label{bias-i-am-inevitable}}

Completely eliminating bias from large language models is a complex and challenging task due to the inherent nature of language and cultural norms. Since these models learn from vast amounts of text data available on the internet, they are exposed to the biases present within human language and culture. Addressing bias in these models involves tackling several key challenges (\textit{cf.}, Table \ref{table:challenges}).

First, human language is a reflection of society and as such, it contains various biases, stereotypes, and assumptions. Separating useful patterns from these biases can be challenging, as they are often deeply ingrained in the way people express themselves and the structures of language itself (\cite{bourdieu1991language, fairclough2001language, lakoff2008metaphors, hill2009everyday, whorf2012language, foucault2013archaeology}).
Second, cultural norms and values can vary significantly across different communities and regions. What is considered acceptable or appropriate in one context may be seen as biased or harmful in another. Determining which norms should be encoded in AI models and which should be filtered out is a complex task that requires careful consideration and a nuanced understanding of diverse cultural perspectives (\cite{geertz1973interpretation, hofstede1984culture, inglehart2005christian, triandis2018individualism}).

Furthermore, fairness is a subjective concept that can be interpreted in various ways. Completely eliminating bias from AI models would require developers to define what ``fair'' means in the context of their applications, which can be a challenging task, given the diverse range of stakeholders and perspectives involved (\cite{friedler2016possibility, zafar2017fairness, barocas2019fairness}).
Lastly, language and culture are constantly evolving, with new expressions, norms, and biases emerging over time. Keeping AI models up-to-date with these changes and ensuring that they remain unbiased is an ongoing challenge that requires continuous monitoring and adaptation (\cite{mufwene2001ecology, jenkins2008convergence, castells2011rise}).

Despite these challenges, it is essential for developers, researchers, and stakeholders to continue working towards reducing bias in large language models. By developing strategies for identifying and mitigating biases, collaborating with diverse communities, and engaging in ongoing evaluation and improvement, we can strive to create AI systems that are more equitable, fair, and beneficial for all users.

\begin{table}[t!]
\centering\footnotesize
\begin{tabular}{|p{3cm}|p{8cm}|p{4cm}|}
\hline
\textbf{Challenge} & \textbf{Description} & \textbf{References} \\
\hline
Inherent biases in language & Human language is a reflection of society, containing various biases, stereotypes, and assumptions. Separating useful patterns from these biases can be challenging as they are deeply ingrained in language structures and expressions. & \cite{bourdieu1991language, fairclough2001language, lakoff2008metaphors, hill2009everyday, whorf2012language, foucault2013archaeology} \\
\hline
Ambiguity of cultural norms & Cultural norms and values vary significantly across communities and regions. Determining which norms to encode in AI models is a complex task that requires a nuanced understanding of diverse cultural perspectives. & \cite{geertz1973interpretation, hofstede1984culture, inglehart2005christian, triandis2018individualism} \\
\hline
Subjectivity of fairness & Fairness is a subjective concept with various interpretations. Eliminating bias from AI models requires defining ``fair'' in the context of applications, which is challenging due to the diverse range of stakeholders and perspectives. & \cite{friedler2016possibility, zafar2017fairness, barocas2019fairness} \\
\hline
Continuously evolving language and culture & Language and culture constantly evolve, with new expressions, norms, and biases emerging over time. Keeping AI models up-to-date with these changes and ensuring they remain unbiased requires continuous monitoring and adaptation. & \cite{mufwene2001ecology, jenkins2008convergence, castells2011rise} \\
\hline
\end{tabular}
\caption{Challenges in addressing biases in Large Language Models}
\label{table:challenges}
\end{table}

\hypertarget{utility-despite-bias}{%
\subsection{Utility despite bias?}\label{utility-despite-bias}}

Biased AI models can
still be useful in certain contexts or applications, as long as users
are aware of their limitations and take them into account when making
decisions. In some cases, the biases present in these models may even be
representative of the real-world context in which they are being used,
providing valuable insights in surfacing societal inequalities that need to be tackled at their root.

The key to leveraging biased AI models responsibly is to ensure that
users have a clear understanding of the potential biases and limitations
associated with these models, so they can make informed decisions about
whether and how to use them in different contexts. Some strategies for
addressing this issue include:

\begin{itemize}
    \item 
\textbf{Transparency}: Developers should be transparent about the methodologies,
data sources, and potential biases of their AI models, providing users
with the necessary information to understand the factors that may
influence the model's predictions and decisions. Best practices about the documentation of models and data have been advanced by the AI community (\cite{modelsheets, datasheets}).

    \item
\textbf{Education and awareness}: Providing resources, training, and support to
help users better understand the potential biases in AI models and how
to account for them when making decisions. This may involve creating
guidelines, best practices, or other educational materials that explain the
implications of bias in AI and how to navigate it responsibly.
    \item
\textbf{Context-specific applications}: In some limited cases, biased AI models may be
viable for specific applications or contexts where their biases
align with the relevant factors or considerations. Experts should be employed
to carefully evaluate the appropriateness of using biased models in these
situations, taking into account the potential risks and benefits
associated with their use, and actionable plans to recognize, quantify, and mitigate biases.
    \item
\textbf{Continuous monitoring and evaluation}: Regularly assessing the
performance of AI models in real-world contexts, monitoring their impact
on users and affected communities, and making adjustments as needed to
address any biases or unintended consequences that emerge over time (\cite{parrots}).

\end{itemize}

By acknowledging that biased models can still be useful in limited
contexts and taking steps to ensure that users are aware of their
limitations and qualified to recognize and mitigate them, we can promote the responsible use of AI technologies and
harness their potential benefits while minimizing the risks associated
with bias.

\hypertarget{eliminating-bias}{%
\section{The broader risks of generative AI bias}\label{eliminating-bias}}

\subsection{Pillars of responsible generative AI development}

Ethical considerations of fairness and equality play a crucial role in the development and deployment of generative AI applications. As these models integrate increasingly into various aspects of our lives, their potential impact on individuals and society as a whole becomes a matter of significant concern. The responsibility lies with developers, researchers, and stakeholders to ensure that AI models treat all users and groups equitably, avoiding the perpetuation of existing biases or the creation of new ones (\textit{cf.}, Table \ref{table:responsible_ai_pillars}).

One key ethical consideration is \textit{representation}. It is essential to ensure that the training data used to develop generative AI models are representative of the diverse range of perspectives, experiences, and backgrounds that exist within society (\cite{torralba2011unbiased, buolamwini2018gender, barocas2019fairness, modelsheets, sun2019mitigating}). This helps to reduce the risk that biases are absorbed and propagated by models, leading to more equitable outcomes.
\textit{Transparency} is another important aspect. Developers should be transparent about the methodologies, data sources, and potential limitations of their generative AI models (\cite{larsson2020transparency, ehsan2021expanding}). This enables users to better understand the factors that may influence the model's predictions and decisions.
\textit{Accountability} is also crucial for responsible generative AI development. Developers and stakeholders must establish a clear framework for accountability, which may include monitoring the performance of AI models, addressing biases and errors, and responding to users and communities affected (\cite{wachter2017transparent, raji2020closing, smith2021clinical}). An unique aspect of generative AI, compared to traditional machine learning, is its ability to possibly replace human artistic expression or to plagiarize the style and uniqueness of human work: as such, preservation of intellectual property, copyright protection and prevention of plagiarism are paramount \cite{ferrara2023genai}.
\textit{Inclusivity} is another key ethical consideration. Generative AI applications should be designed to be inclusive and accessible to all users, taking into account factors such as language, culture, and accessibility needs (\cite{morris2020ai, schwartz2020green}). This ensures that the benefits of AI are shared equitably across society.

Lastly, \textit{continuous improvement} is vital to achieve fairness and equality in generative AI applications. Developers must commit to an ongoing process of evaluating, refining, and improving their AI models to address biases and ensure fairness over time (\cite{o2017weapons, holstein2019improving, modelsheets, challen2019artificial}). This may involve collaborating with researchers, policymakers, and affected communities to gain insight and feedback that can help guide the development of more equitable AI systems.

By prioritizing ethical considerations of fairness and equality, AI developers can create applications that not only harness the power of advanced technologies such as large language models, but also promote a more just and inclusive society, where the benefits and opportunities of AI are accessible to all.

\begin{table}[t!]
\centering\footnotesize
\begin{tabular}{|p{2.2cm}|p{9cm}|p{4cm}|}
\hline
\textbf{Pillar} & \textbf{Description} & \textbf{References} \\
\hline
Representation & Ensuring that the training data used to develop AI models is representative of the diverse range of perspectives, experiences, and backgrounds that exist within society. This helps to reduce the risk of biases being absorbed and propagated by the models, leading to more equitable outcomes & \cite{torralba2011unbiased, buolamwini2018gender, barocas2019fairness, modelsheets, sun2019mitigating} \\
\hline
Transparency & Developers should be transparent about the methodologies, data sources, and potential limitations of their AI models, enabling users to better understand the factors that may influence the model's predictions and decisions & \cite{larsson2020transparency, ehsan2021expanding} \\
\hline
Accountability & It is essential for developers and stakeholders to establish a clear framework for accountability, which may include monitoring the performance of AI models, addressing biases and errors, and responding to the concerns of users and affected communities & 
\cite{wachter2017transparent, raji2020closing, smith2021clinical} \\
\hline
Inclusivity & AI applications should be designed to be inclusive and accessible to all users, taking into account factors such as language, culture, and accessibility needs, to ensure that the benefits of AI are shared equitably across society & 
\cite{morris2020ai, schwartz2020green} \\
\hline
Protection of IP, human work, and human artistic expression & Generative AI models have the remarkable capability to create human-like text, artwork, music, and more. This creative aspect presents unique challenges, including issues related to intellectual property and the protection of human-generated copyright work to avoid AI plagiarism & \cite{ferrara2023genai} \\
\hline
Continuous improvement & Developers must commit to an ongoing process of evaluating, refining, and improving their AI models to address biases and ensure fairness over time. This may involve working with researchers, policy makers, and affected communities to gain information and feedback that can help guide the development of more equitable AI systems & 
\cite{o2017weapons, holstein2019improving, modelsheets, challen2019artificial} \\
\hline
\end{tabular}
\caption{Pillars of responsible generative AI development}
\label{table:responsible_ai_pillars}
\end{table}

\hypertarget{exacerbating-existing-societal-biases}{%
\subsection{The risks of exacerbating existing societal
biases}\label{exacerbating-existing-societal-biases}}

Bias in widely-adopted AI models, including ChatGPT and other generative language models,
can have far-reaching consequences that extend beyond the immediate
context of their applications. When these models absorb and propagate
biases, including those present in their training data, they may inadvertently reinforce
stereotypes, marginalize certain groups, and lead to unfair treatment
across various domains. Some examples of how biased AI models can
adversely impact different areas include:

\begin{itemize}

\item
  \textbf{Hiring}: AI-driven hiring tools that use biased models may exhibit
  unfair treatment towards applicants from underrepresented groups or
  those with non-traditional backgrounds. This could lead to the
  perpetuation of existing inequalities in the job market, limiting
  opportunities for affected individuals and reducing diversity in the
  workforce (\cite{bogen2018help, raghavan2020mitigating}).
  Large language models can be used to automate the screening of job applicants, such as by analyzing resumes and cover letters. Since these models are trained on vast amounts of text data, they may have internalized biases present in the data, such as gender or racial biases. As a result, they could unintentionally favor certain applicants or disqualify others based on factors unrelated to their qualifications, reinforcing existing inequalities in the job market.

\item
  \textbf{Lending}: Financial institutions increasingly rely on AI models for
  credit scoring and lending decisions. Biased models may unfairly
  penalize certain groups, such as minority communities or individuals
  with lower socio-economic status, by assigning them lower credit
  scores or denying them access to loans and financial services based on
  biased assumptions (\cite{citron2014scored, lee2019algorithmic, ustun2019actionable}).
  In lending, large language models can be used to assess creditworthiness or predict loan default risk, e.g., based on automated analysis of application or support documents. If the data used to train these models contain historical biases or discriminatory lending practices, the models may learn to replicate these patterns. Consequently, they could deny loans to certain demographics or offer unfavorable terms based on factors like race, gender, or socioeconomic status, perpetuating financial inequality (\cite{weidinger2021ethical}).

\item
  \textbf{Content moderation}: AI-powered content moderation systems help manage
  and filter user-generated content on social media platforms and other
  online communities. If these systems are trained on biased data, they
  may disproportionately censor or suppress the voices of certain
  groups, while allowing harmful content or misinformation from other
  sources to proliferate (\cite{gillespie2018custodians, roberts2019behind, choi2023automated, augenstein2023factuality}).
  Language models can be employed to automatically moderate and filter content on social media platforms or online forums. However, these models may struggle to understand the nuances of language, context, and cultural differences. They might over-moderate or under-moderate certain types of content, disproportionately affecting certain groups or topics. This could lead to censorship or the amplification of harmful content, perpetuating biases and misinformation (\cite{sap2019risk, davidson2019racial, pasquetto2020tackling, ezzeddine2023exposing}).

\item
  \textbf{Healthcare}: AI models are increasingly used to support medical
  decision-making and resource allocation. Biased models may result in
  unfair treatment for certain patient groups, leading to disparities in
  healthcare access and quality, and potentially exacerbating existing
  health inequalities (\cite{gianfrancesco2018potential, obermeyer2019dissecting, challen2019artificial, smith2021clinical}).
  Large language models can be employed for tasks such as diagnosing diseases, recommending treatments, or analyzing patient data (\cite{davenport2019potential, ngiam2019big}). If the training data includes biased or unrepresentative information, the models may produce biased outcomes. Data used to train these models might be predominantly collected from specific populations, leading to less accurate predictions or recommendations for underrepresented groups. This can result in misdiagnoses, inadequate treatment plans, or unequal access to care. Models might unintentionally learn to associate certain diseases or conditions with specific demographic factors, perpetuating stereotypes and potentially influencing healthcare professionals' decision-making. Finally, biases in healthcare data could lead to models that prioritize certain types of treatments or interventions over others, disproportionately benefiting certain groups and disadvantaging others (\cite{paulus2020predictably, ferrara2023fairness}).

\item
  \textbf{Education}: AI-driven educational tools and platforms can help
  personalize learning experiences and improve educational outcomes.
  However, biased models may perpetuate disparities by favoring certain
  learning styles or cultural backgrounds, disadvantaging students from
  underrepresented or marginalized communities (\cite{elish2018situating, selwyn2019should}). Large language models can be used in education for tasks such as personalized learning, grading, or content creation. If the models are trained on biased data, they may exacerbate existing biases in educational settings. Furthermore, if used for grading or assessing student work, language models might internalize biases in historical grading practices, leading to unfair evaluation of students based on factors like race, gender, or socioeconomic status. Finally, access in itself to ChatGPT or other AI tools for education can exacerbate preexisting inequalities.

\end{itemize}

\subsection{Paths to AI transparency}
Transparency and trust are essential components in the development and
deployment of AI systems. As these models become more integrated into various aspects of
our lives, it is increasingly important for users and regulators to
understand how they make decisions and predictions, ensuring that they
operate fairly, ethically, and responsibly.

Emphasizing transparency in AI systems can provide several benefits:

\begin{itemize}

\item
  \textbf{Informed decision-making}: When users and regulators have a clear
  understanding of how AI models make decisions and predictions, they
  can make more informed choices about whether to use or rely on these
  systems in different contexts. Transparency can empower users to
  evaluate the potential risks and benefits of AI systems and make
  decisions that align with their values and priorities (\cite{goodman2017european, guidotti2018survey}).
\item
  \textbf{Public trust}: Fostering transparency can help build public trust in AI
  systems, as it demonstrates a commitment to ethical development and
  responsible deployment. When users and regulators can see that
  developers are taking steps to ensure the fairness and equity of their
  models, they may be more likely to trust and adopt these technologies (\cite{larsson2020transparency, ehsan2021expanding}).
\item
  \textbf{Ethical compliance}: By promoting transparency, developers can
  demonstrate their compliance with ethical guidelines and regulations,
  showcasing their commitment to the responsible development of AI
  systems. This can help to establish a strong reputation and foster
  positive relationships with users, regulators, and other stakeholders (\cite{jobin2019global, weidinger2021ethical}).
\item
  \textbf{Collaborative improvement}: Transparency can facilitate collaboration
  between developers, researchers, policymakers, and affected
  communities, enabling them to share insights and feedback that can
  help guide the development of more equitable and ethical AI systems (\cite{o2017weapons, holstein2019improving, modelsheets, challen2019artificial}).
\end{itemize}

In summary, emphasizing transparency and trust in AI systems, including generative language models, is crucial for ensuring that users and regulators have a clear understanding of how these models make decisions and predictions. By promoting transparency, developers can demonstrate their commitment to ethical development and responsible deployment, fostering public trust and paving the way for more equitable and beneficial AI applications.

\hypertarget{regulatory-efforts-industry-standards-and-ethical-guidelines}{%
\subsection{Regulatory efforts, industry standards, and ethical
guidelines}\label{regulatory-efforts-industry-standards-and-ethical-guidelines}}

As concerns about bias in AI systems continue to grow, several ongoing
regulatory efforts and industry standards have emerged to address these
challenges. AI ethics guidelines and fairness frameworks aim to provide
guidance and best practices for developers, organizations, and
policymakers to reduce bias and ensure responsible development and
deployment of AI systems (\cite{jobin2019global, floridi2019establishing}). 

Some notable efforts are summarized in Table \ref{tab:regulatory_efforts}.
These ongoing regulatory efforts and industry standards represent
important steps in addressing bias and promoting the responsible
development of AI systems. By adhering to these guidelines and
frameworks, developers can contribute to the creation of AI technologies
that are more equitable, fair, and beneficial for all users and
communities.

\begin{table}[t!]
\centering\footnotesize
\caption{Ongoing Regulatory Efforts and Industry Standards to Address Bias in AI}
\label{tab:regulatory_efforts}
\begin{tabularx}{\textwidth}{p{3cm}X}
\toprule
\textbf{Effort} & \textbf{Description} \\
\bottomrule
European Union's AI Ethics Guidelines & The EU's High-Level Expert Group on Artificial Intelligence has developed a set of AI Ethics Guidelines that outline key ethical principles and requirements for trustworthy AI, including fairness, transparency, accountability, and human agency (\cite{EUAIethics}). \\
\addlinespace
IEEE's Ethically Aligned Design & The Institute of Electrical and Electronics Engineers (IEEE) has published a comprehensive document, "Ethically Aligned Design," which provides recommendations and guidelines for the ethical development of AI and autonomous systems, with a focus on human rights, data agency, and technical robustness (\cite{shahriari2017ieee}). \\
\addlinespace
Partnership on AI & A coalition of tech companies, research institutes, and civil society organizations, the Partnership on AI aims to promote the responsible development and use of AI technologies. They work on various initiatives, including fairness, transparency, and accountability, to ensure that AI benefits all of humanity (\cite{heer2018partnership}). \\
\addlinespace
AI Fairness 360 (AIF360) & Developed by IBM Research, AIF360 is an open-source toolkit that provides a comprehensive suite of metrics and algorithms to help developers detect and mitigate bias in their AI models. It assists developers in understanding and addressing fairness concerns in their AI applications (\cite{bellamy2019ai}). \\
\addlinespace
Google's AI Principles & Google has outlined a set of AI Principles that guide the ethical development and use of AI technologies within the company. These principles emphasize fairness, transparency, accountability, and the importance of avoiding harmful or unjust impacts (\cite{pichai2018ai}). \\
\addlinespace
Algorithmic Impact Assessment (AIA) & Developed by the AI Now Institute, the AIA is a framework designed to help organizations evaluate and mitigate the potential risks and harms of AI systems. The AIA guides organizations through a structured process of identifying and addressing potential biases, discrimination, and other negative consequences of AI deployment (\cite{reisman2018algorithmic}). \\
\addlinespace
OECD Recommendation of the Council on Artificial Intelligence & The OECD Recommendation of the Council on Artificial Intelligence is a set of guidelines that provides a framework for the responsible development and deployment of AI, with five principles focused on inclusive growth, human-centered values, transparency and explainability, robustness, security and safety, and accountability (\cite{cath2018artificial, yeung2020recommendation}). \\
\bottomrule
\end{tabularx}
\end{table}

\hypertarget{the-role-of-human-oversight-and-intervention}{%
\section{The role of human oversight and
intervention}\label{the-role-of-human-oversight-and-intervention}}

\hypertarget{how-to-identify-and-mitigate-bias}{%
\subsection{How to identify and mitigate
bias?}\label{how-to-identify-and-mitigate-bias}}

Identifying and mitigating bias in AI models is essential for ensuring
their responsible and equitable use. Various methods can be employed to
address bias in AI systems:

\begin{itemize}
    \item 

\textbf{Regular audits}: Conducting regular audits of AI models can help identify
potential biases, errors, or unintended consequences in their outputs.
These audits involve evaluating the model's performance against a set of
predefined metrics and criteria, which may include fairness, accuracy,
and representativeness. By monitoring AI models on an ongoing basis,
developers can detect and address biases before they become problematic (\cite{raji2020closing}).
    \item 
\textbf{Retraining with curated data}: Retraining AI models with curated data can
help reduce biases in their predictions and decisions. By carefully
selecting and preparing training data that is more diverse, balanced,
and representative of different perspectives, developers can ensure that
AI models learn from a broader range of inputs and experiences, which
may help mitigate the influence of biases present in the original
training data (\cite{gururangan2020don, instructgpt}).
    \item 
\textbf{Applying fairness metrics}: Fairness metrics can be used to evaluate the
performance of AI models with respect to different user groups or
populations (\cite{yan2020fair, yang2023fairfed}). By analyzing AI model outputs based on these metrics,
developers can identify potential disparities or biases in the model's
treatment of different users and take steps to address them. Examples of
fairness metrics include demographic parity, equalized odds, and equal
opportunity (\cite{mehrabi2021survey}).
    \item 
\textbf{Algorithmic debiasing techniques}: Various algorithmic techniques have
been developed to mitigate bias in AI models during training or
post-processing. Some of these techniques include adversarial training,
re-sampling, and re-weighting, which aim to minimize the influence of
biased patterns and features on the model's predictions and decisions (\cite{bolukbasi2016man, zhang2018mitigating, bender2018data, dev2019attenuating, bordia2019identifying, sun2019mitigating, lee2019algorithmic, raghavan2020mitigating}).
    \item 
\textbf{Inclusion of diverse perspectives}: Ensuring that AI development teams
are diverse and inclusive can help bring a wide range of perspectives
and experiences to the table, which can contribute to the identification
and mitigation of biases in AI models. By involving individuals from
different backgrounds, cultures, and disciplines, developers can create
more robust, fair, and representative AI systems (\cite{piledataset, piledatasheet}).
    \item 
\textbf{Human-in-the-loop approaches}: Incorporating human experts into the AI
model development and decision-making processes can help provide
valuable contextual understanding and ethical judgment that AI models
may lack. Humans can serve as an additional layer of quality control,
identifying biases, errors, or unintended consequences in AI model
outputs and providing feedback to improve the model's performance and
fairness.

\end{itemize}

Next, we dive deeper into the role that human experts can play in the responsible design and development of AI systems, including large language models, and their continuous oversight.

\subsection{The importance of humans in the AI loop}
Emphasizing the importance of involving human experts in AI system
development, monitoring, and decision-making is crucial to ensuring the
responsible and ethical deployment of AI technologies. Humans possess the ability to provide
context and ethical judgment that AI models may lack, and their
involvement can help address potential biases, errors, and unintended
consequences that may arise from the use of these systems.
Some key benefits of involving human experts in AI system development
include:

\begin{itemize}
    \item 

\textbf{Contextual understanding}: Human experts can provide valuable insights
into the cultural, social, and historical contexts that shape language
and communication, helping to guide AI models in generating more
appropriate and sensitive responses (\cite{elish2018situating, leslie2019understanding, dwivedi2021artificial}).
    \item 

\textbf{Ethical judgment}: Human experts possess the moral and ethical reasoning
skills needed to evaluate the potential impacts of AI systems on users
and affected communities. By involving human experts in decision-making,
we can ensure that AI models align with ethical principles and values,
such as fairness, transparency, and accountability (\cite{shahriari2017ieee, jobin2019global, smuha2019eu, EUAIethics, weidinger2021ethical}).
    \item 

\textbf{Bias identification and mitigation}: Human experts can help identify and
address biases in AI models, working alongside developers to implement
strategies for mitigating or eliminating harmful biases and promoting
more equitable and representative AI systems (\cite{bender2018data, sun2019mitigating, raghavan2020mitigating}).
    \item 

\textbf{Quality assurance and validation}: Human experts can serve as a vital
layer of quality control, evaluating AI model outputs for coherence,
relevance, and potential biases, and providing feedback to improve the
model's performance, accuracy, regulatory compliance, and trustworthiness  (\cite{felderer2021quality}).
    \item 

\textbf{Human override}: Incorporating human experts into AI system
workflows can help strike a balance between automation and human
judgment, allowing humans to intervene and override AI model decisions
when necessary to ensure fairness, accountability, and ethical compliance (\cite{etzioni2016keeping}).

\end{itemize}

By involving human experts in the development, monitoring, and
decision-making processes of AI systems, we can leverage their
contextual understanding and ethical judgment to complement the
capabilities of AI models. This collaborative approach can help us
create AI systems that are more responsible, equitable, and beneficial
for all users, while also addressing the potential risks and challenges
associated with bias in AI.

\begin{table}[t]
\centering\footnotesize
\begin{tabular}{|p{3cm}|p{7.5cm}|p{4.5cm}|}
\hline
\textbf{Strategy} & \textbf{Description} & \textbf{References} \\ \hline
Engaging with affected communities & Involving affected communities in the development and evaluation of AI models can lead to the creation of generative AI systems that are more culturally sensitive, contextually relevant, and fair to all users & \cite{wallach2008machine, taylor2015bigger, mittelstadt2016ethics, eubanks2018automating, gray2019ghost, west2019discriminating, costanza2020design} \\ \hline
Multidisciplinary collaboration & Bringing together experts from different fields, such as computer science, social sciences, humanities, and ethics, can help to develop more robust strategies for addressing and mitigating bias in generative AI systems & \cite{holstein2019improving, crawford2019ai, mittelstadt2016ethics} \\ \hline
User feedback and evaluation & Encouraging users to provide feedback on AI model outputs and performance can contribute to the ongoing improvement and refinement of generative AI models, ensuring that they remain fair, accurate, and relevant to users' needs & \cite{cramer2008effects, lee2015working, amershi2019guidelines, stoyanovich2020responsible} \\ \hline
Openness and transparency & Sharing information about the methodologies, data sources, and potential biases of generative AI models can enable stakeholders to make more informed decisions about whether and how to use these technologies in different contexts & \cite{mittelstadt2016ethics, taddeo2018regulate, cath2018artificial, jobin2019global} \\ \hline
Establishing partnerships & Forming partnerships between AI developers, research institutions, non-profit organizations, and other stakeholders can facilitate the sharing of knowledge, resources, and best practices, leading to the development of more equitable and responsible AI technologies & \cite{jobin2019global} \\ \hline
\end{tabular}
\caption{Strategies for addressing bias in generative AI systems}
\label{tab:strategies}
\end{table}

\hypertarget{possible-strategies-and-best-practices-to-address-bias}{%
\subsection{Possible strategies and best practices to address bias in generative AI} \label{possible-strategies-and-best-practices-to-address-bias}}

Addressing and mitigating potential biases in generative AI models requires a collaborative effort between
AI developers, users, and affected communities. Fostering a more
inclusive and fair AI ecosystem involves engaging various stakeholders
in the development, evaluation, and deployment of AI technologies. This
collaboration can oversee that AI models are designed to be more
equitable, representative, and beneficial to all users.
 
Some key aspects of fostering collaboration in the AI ecosystem are shown in Table \ref{tab:strategies}.
By fostering collaboration between AI developers, users, and affected
communities, we can work towards creating a more inclusive and fair AI
ecosystem that respects and values diverse perspectives and experiences.
This collaborative approach can help ensure that AI technologies are
developed and deployed in a way that is equitable, responsible, and
beneficial for all users, while also addressing the potential risks and
challenges associated with bias in AI.

\begin{table}[t!]
\centering\footnotesize
\caption{Future avenues for Large Language Models and generative AI research.}
\label{tab:future_research_avenues}
\begin{tabularx}{\textwidth}{lX}
\toprule
\textbf{Research Area} & \textbf{Description} \\
\midrule
Fairness, Bias, and Ethics & Addressing and minimizing biases in language models, as well as understanding their ethical implications, is a critical area of research; developing methods to detect, mitigate and prevent biases in AI models is essential. \\
\addlinespace
Interpretability and Explainability & Understanding the internal workings of large language models is a significant challenge; researchers are developing methods to make models more interpretable and explain their predictions. \\
\addlinespace
Auditability and Accountability & Large language models increasingly impact various sectors of society, influencing decision-making and shaping public discourse; ensuring that models are transparent, their actions can be traced, and those responsible for their development and deployment can be held accountable for the consequences of the AI's actions is vital for fostering trust and maintaining ethical standards in the AI community and beyond. \\
\addlinespace
Controllable and Safe AI & Ensuring that AI models can generate outputs that align with human intentions and values is an important research question; developing methods to control AI behavior, reduce harmful outputs and improve safety measures is vital. \\
\addlinespace
Societal Effects & The societal effects and implications of the deployment of AI systems encompass a wide range of concerns, including labor markets, privacy, bias, access to technology, public discourse, security, ethics, and regulation; Observing, characterizing, quantifying and understanding the broader effects that the deployment of AI systems has on society warrant careful consideration and continued research as these technologies continue to proliferate. \\
\bottomrule
\end{tabularx}
\end{table}

\hypertarget{conclusions}{%
\section{Conclusions}\label{conclusions}}

This paper highlights the challenges and risks associated with biases in generative language models like ChatGPT, emphasizing the need for a multi-disciplinary, collaborative effort to develop more equitable, transparent, and responsible AI systems that enhance a wide array of applications while minimizing unintended consequences.

Various methods for identifying and mitigating bias in AI models were presented, including regular audits, retraining with curated data, applying fairness metrics, and incorporating human experts in AI system development, monitoring, and decision-making.
To achieve this goal, the development and deployment of AI technologies should prioritize ethical principles, such as fairness and equality, ensuring that all users and groups are treated equitably. Human oversight plays a vital role in providing context and ethical judgment that AI models may lack, helping to identify and address potential biases, errors, or unintended consequences. Collaboration between AI developers, users, and affected communities is essential for fostering a more inclusive and fair AI ecosystem, ensuring that diverse perspectives and experiences are considered and valued.

Continued research into methods for identifying, addressing, and mitigating biases in AI models will be critical to advancing the state of the art and promoting more equitable and inclusive AI systems. By bringing together experts from various disciplines, including computer science, social sciences, humanities, and ethics, we can foster a more comprehensive understanding of the potential biases and ethical challenges associated with AI applications.

Fostering an open and ongoing dialogue between stakeholders is crucial for sharing knowledge, best practices, and lessons learned from the development and use of AI applications. This dialogue can help to raise awareness of the potential risks and challenges associated with biases in AI models and promote the development of strategies and guidelines for mitigating their negative impacts.

\paragraph{Future research avenues}

As the development of large language models continues, several essential aspects of research are necessary to advance their understanding and ensure their responsible deployment (\textit{cf.}, Table \ref{tab:future_research_avenues}). Among these are understanding their inner workings, addressing ethical concerns, ensuring controllability and safety, and developing more robust evaluation methods. 

One critical area of research is fairness, bias, and ethics, which involves detecting, mitigating, and preventing biases in AI models to minimize their impact on various sectors of society. Another important research question is interpretability and explainability, as understanding the internal workings of large language models remains a significant challenge. Researchers are working to make models more interpretable and explain their predictions. Additionally, large language models can influence decision-making and shape public discourse, making auditability and accountability essential for fostering trust and maintaining ethical standards in the AI community and beyond. 

Controllable and safe AI is also important to ensure that AI models can generate outputs that align with human intentions and values. Finally, observing, characterizing, quantifying, and understanding the broader effects that deploying AI systems has on society is critical for addressing societal effects, which encompass a wide range of concerns, including labor markets, privacy, bias, access to technology, public discourse, security, ethics, and regulation. The need for a multi-disciplinary, collaborative effort to develop more equitable, transparent, and responsible AI systems is clear. 

This paper emphasized the challenges and risks associated with biases in generative language models like ChatGPT and highlights the importance of continued research to develop responsible AI systems that enhance a wide array of applications while minimizing unintended consequences.

\section*{Acknowledgements}
The author is grateful to all current and past members of his lab at USC, and the numerous colleagues and students at USC Viterbi and Annenberg, who engaged in stimulating discussions and provided invaluable feedback about this study.
\bibliography{chatgpt}

\bibliographystyle{elsarticle-num} 





\end{document}